\documentclass{sig-alternate}

\usepackage{amsmath}
\usepackage{graphicx}
\usepackage{listings}
\usepackage{subfigure}
\usepackage{verbatim}
\usepackage{parcolumns}
\usepackage{courier}

\lstdefinelanguage
[x64]{Assembler}     
[x86masm]{Assembler} 
{morekeywords={
  CDQE,CQO,CMPSQ,CMPXCHG16B,JRCXZ,LODSQ,MOVSXD,CMPL,NEGL,CMOVEL,
  POPFQ,PUSHFQ,SCASQ,STOSQ,IRETQ,RDTSCP,SWAPGS, 
  MOVQ,SHRQ,ANDL,IMULQ,ADDQ,MOVABSQ,ADDQ,SALQ,ADCQ,XORQ,MULQ,SHLQ, 
  LEAL,TESTL,XORL,POPCNTL,DECL,MOVL,SARL,IMULL,ADDL,SUBL,
  SHRL,MOVZWL,MOVSLQ,MOVD,SHUFPS,PMULLw,PADDw,LEAQ,XMM0,XMM1,MOVAPS,SALL,TESTQ,MOVUPS,
  r10,r10d,.SET,
  rax,rdx,rcx,rbx,rsi,rdi,rsp,rbp, 
  r8,r8d,r8w,r8b,r9,r9d,r9w,r9b,
  bool,void,return,uint,while,struct}} 

\lstnewenvironment{acol}[1][]{\lstset{language=[x64]Assembler, basicstyle=\footnotesize}}{}

\begin{document}
%
% --- Author Metadata here ---
\conferenceinfo{ASPLOS}{'13 Houston, Texas USA}
%\CopyrightYear{2007} % Allows default copyright year (20XX) to be over-ridden - IF NEED BE.
%\crdata{0-12345-67-8/90/01}  % Allows default copyright data (0-89791-88-6/97/05) to be over-ridden - IF NEED BE.
% --- End of Author Metadata ---

\title{Stochastic Superoptimization}

\numberofauthors{3} %  in this sample file, there are a *total*

\author{
% 1st. author
\alignauthor Eric Schkufza \\
       \affaddr{Stanford University}\\
       \affaddr{Stanford, CA}\\
       \email{eschkufz@cs.stanford.edu}
% 2nd. author
\alignauthor Rahul Sharma \\
       \affaddr{Stanford University}\\
       \affaddr{Stanford, CA}\\
       \email{sharmar@cs.stanford.edu}
% 3rd. author
\alignauthor Alex Aiken \\
       \affaddr{Stanford University}\\
       \affaddr{Stanford, CA}\\
       \email{aiken@cs.stanford.edu}
}

% There's nothing stopping you putting the seventh, eighth, etc.
% author on the opening page (as the 'third row') but we ask,
% for aesthetic reasons that you place these 'additional authors'
% in the \additional authors block, viz.
%\additionalauthors{Additional authors: John Smith (The Th{\o}rv{\"a}ld Group,
%email: {\texttt{jsmith@affiliation.org}}) and Julius P.~Kumquat
%(The Kumquat Consortium, email: {\texttt{jpkumquat@consortium.net}}).}
\date{23 July 2012}
% Just remember to make sure that the TOTAL number of authors
% is the number that will appear on the first page PLUS the
% number that will appear in the \additionalauthors section.

\maketitle

\begin{abstract}
We formulate the loop-free, binary superoptimization task as a stochastic search problem.
The competing constraints of transformation correctness and performance improvement are encoded as terms in a cost function, 
  and a Markov Chain Monte Carlo sampler is used to rapidly explore the space of all possible programs to find one that is
an optimization of a given target program.
Although our method sacrifices completeness, the scope of programs we are able to reason about, 
  and the quality of the programs we produce, far exceed those of existing superoptimizers.
Beginning from binaries compiled by llvm -O0 for 64-bit X86, our prototype implementation, STOKE, 
  is able to produce programs which either match or outperform the code sequences produced by gcc with full optimizations enabled, and, in some cases, expert handwritten assembly. 
\end{abstract}

%A category including the fourth, optional field follows...



\section{Introduction}

\begin{figure}[h]
    \tt
\begin{lstlisting}
# rsi=np, ecx=mh, edx=ml, rdi=c0, r8=c1
# c1 : c0 := np * mh:ml + c1 + c0

.set c0 0xffffffff
.set c1 0x100000000

    \end{lstlisting}
    \begin{parcolumns}[rulebetween=true]{2} \colchunk[1]{\begin{acol}
.L0                  
  movq    rsi,  r9   
  mov     ecx,  ecx  
  shrq    32,   rsi  
  andl    c1,   r9d  
  movq    rcx,  rax  
  mov     edx,  edx  
  imulq   r9,   rax  
  imulq   rdx,  r9   
  imulq   rsi,  rdx  
  imulq   rsi,  rcx  
  addq    rdx,  rax  
  jae     .L2        
  movabsq c1,   rdx
  addq    rdx,  rcx
.L2
  movq    rax,  rsi
  movq    rax,  rdx
  shrq    32,   rsi
  salq    32,   rdx
  addq    rsi,  rcx
  addq    r9,   rdx
  adcq    0,    rcx
  addq    r8,   rdx
  adcq    0,    rcx
  addq    rdi,  rdx
  adcq    0,    rcx
  movq    rcx,  r8
  movq    rdx,  rdi
\end{acol}}
\colchunk[2]{\begin{acol}
.L0                  
  shlq  32,   rcx
  mov   edx,  edx
  xorq  rdx,  rcx
  movq  rcx,  rax
  mulq  rsi
  addq  r8,   rdi
  adcq  0,    rdx
  addq  rdi,  rax
  adcq  0,    rdx
  movq  rdx,  r8
  movq  rax,  rdi  
\end{acol}}
\end{parcolumns}
\rm
  \centering
  \caption{Montgomery multiplication kernel from the OpenSSL big number library, 
           compiled by gcc -O3 (left) and STOKE (right).
           The STOKE code is 16 lines shorter, 1.6x faster, and slightly faster than expert handwritten assembly.}
  \label{fig:teaser}
\end{figure}

For many application domains there is considerable value in producing the most performant code possible.
Unfortunately, the traditional structure of a compiler's optimization
phase is often ill-suited to this task.  Attempting to factor the
optimization problem into a collection of small subproblems that can
be solved independently, although suitable for generating consistently
good code, leads to the well-known phase ordering problem.  In many
cases, the best possible code can only be obtained through the
simultaneous consideration of mutually dependent issues such as
instruction selection, register allocation, and target-dependent
optimization.
 
Previous approaches to this problem have focused on the exploration of
all possibilities within some limited class of programs.  In contrast
to a traditional compiler, which uses performance constraints to drive
code generation of a single program, these systems consider multiple programs and
then ask how well they satisfy those constraints.  Solutions range
from the explicit enumeration of a class of programs that can be
formed using a large executable hardware instruction set
\cite{DBLP:conf/asplos/BansalA06} to implicit enumeration through
symbolic theorem proving techniques of programs over some restricted
register transaction language \cite{DBLP:conf/asplos/Massalin87,
DBLP:conf/pldi/JoshiNR02,DBLP:conf/pldi/GulwaniJTV11}.

An attractive feature of these systems is completeness: If a program
exists meeting the desired constraints, that program will be found.
Unfortunately, completeness also places limitations on the space of
programs that can be effectively reasoned about.  
Because of the huge number of programs involved
explicit enumeration-based techniques are limited to programs up to some fixed
length, and currently this bound is well below the threshold at which
many interesting optimizations take place.  Implicit enumeration
techniques can overcome this limitation, but at the cost of
expert-written rules for shrinking the search space.  The resulting
optimizations are as good, but no better, than the quality
of the rules written by an expert.

To overcome these limitations we take a different approach based on
incomplete search.  We show how the competing requirements of
correctness and speed can be defined as terms in a cost function over
the complex search space of all loop-free executable hardware
instruction sequences, and how the program optimization problem can be
formulated as a cost minimization problem.  Although the resulting
search space is highly irregular and not amenable to exact
optimization techniques, we demonstrate that the common approach of employing
a Markov Chain Monte Carlo (MCMC) sampler to explore the function and
produce low-cost samples is sufficient for producing high quality
code sequences.

Although our technique sacrifices completeness by trading systematic
enumeration for stochastic search, we show that we are able to
dramatically increase the space of programs that our system can
reason while simultaneously improving the quality of the code produced.
Consider the example code shown in Figure \ref{fig:teaser}, 
  the Montgomery multiplication kernel from the OpenSSL big number library for arbitrary precision integer arithmetic.
Beginning with a binary compiled by llvm -O0 (116 lines, not shown), 
  we are able to produce a program which is 16 lines shorter and 1.6 times faster 
  than the code produced by gcc with full optimizations enabled.
Most interestingly, the code that our method finds uses a different assembly level algorithm
than the original, and is slightly better than the expert handwritten assembly code included with the OpenSSL repository.
The code is discovered automatically, 
  and is automatically verified to be equivalent to the original llvm -O0 code.
To the best of our knowledge, the code is truly optimal: it is the fastest program for this
function written in the 64-bit X86 instruction set
(the strange looking {\tt mov edx, edx} produces the non-obvious but necessary side effect of zeroing the upper 32 bits of {\tt rdx}).

To summarize, our work makes a number of contributions that
have not previously been demonstrated.  The remainder of
this paper explores each in turn.  Section \ref{sec:rel}
summarizes previous work in superoptimization and discusses its limitations.
Section \ref{sec:theory} presents a mathematical formalism
for transforming the program optimization task into a stochastic cost
minimization problem.  Section \ref{sec:bin} discusses how that theory
is applied in a system for optimizing the runtime performance of
64-bit X86 binaries, and Section \ref{sec:sys} describes our prototype
implementation, STOKE.  Finally, Section \ref{sec:eval} evaluates
STOKE on a set of benchmarks drawn from cryptography, linear algebra, 
and low-level programming, and shows that STOKE is able to
produce code that either matches or outperforms the code produced 
by production compilers.

\section{Related Work}
\label{sec:rel}

Previous approaches to superoptimization have focused on the
exploration of all possibilities within some restricted class of
programs.  Although these systems have been demonstrated to be quite
effective within certain domains, their general applicability
has remained limited.  We discuss these limitations in the context of
the Montgomery multiplication kernel shown in Figure \ref{fig:teaser}.

The high-level organization of the code is as follows: Two 32-bit
values, ${\rm ecx}$ and ${\rm edx}$, are concatenated and then
multiplied by the 64-bit ${\rm rsi}$ to produce a 128-bit value.  Two
64-bit values, ${\rm rdi}$ and ${\rm r8}$ are added to that product,
and the result is written to two registers: the upper half to ${\rm
  r8}$, and the lower half to ${\rm rdi}$.  The primary source of
optimization is best demonstrated by comparison.  The code produced by
gcc -O3, Figure \ref{fig:teaser}(left), performs the 128-bit
multiplication as four 64-bit multiplications and then combines the
results; the rewrite produced by STOKE, Figure
\ref{fig:teaser}(right), uses a hardware intrinsic to perform
the multiplication in a single step.

Massalin's original paper on superoptimization
\cite{DBLP:conf/asplos/Massalin87} describes a system that explicitly
enumerates sequences of code of increasing length and selects the first such code
identical to the input program on a set of testcases.  Massalin reported being able to
optimize instruction sequences of up to length 12, however to do so,
it was necessary to restrict the set of enumerable opcodes to between
10 and 15.  The 11 instruction kernel produced by STOKE in Figure~\ref{fig:teaser}
is found by considering a large subset of the nearly 400 64-bit X86 opcodes, some of which
have as many as 10 variations.  It is unlikely that Massalin's
approach would scale to an instruction set of this magnitude.

Denali \cite{DBLP:conf/pldi/JoshiNR02}, and the more recent Equality
Saturation technique \cite{lerner2009}, attempt to gain scalability
by only considering programs that are known to be equal to the input program.
Candidate programs are explored through successive application of
equality preserving transformation axioms.  Because it is goal-directed 
this approach dramatically improves both the number of primitive instructions
and the length of programs that can be considered, but it also relies heavily
on expert knowledge.  It is unclear whether an expert would know a priori to
encode an equality axiom defining the multiplication transformation
discovered by STOKE.  More generally, it is unlikely that a set of
expert written rules would ever cover the set of all interesting
optimizations.  It is worth noting that these techniques can to a certain 
extent deal with loop optimizations, while other techniques, including ours,
are limited to loop-free code.

Bansal \cite{DBLP:conf/asplos/BansalA06} 
describes a system that automatically enumerates 32-bit X86
superoptimizations and stores the results in a database for later use.
By exploiting symmetries between programs that are equivalent up to
register renaming, Bansal was able to scale this method to
optimizations taking input code sequences of at most length 6 and producing
code sequences of at most length 3.  This approach has the dual benefit of
hiding the high cost of superoptimization by performing a search once
and for all offline and eliminating the dependence on expert
knowledge.  To some extent, the low cost of performing a database
query allows the system to overcome the low upper bound on instruction
length through the repeated application of the optimizer along a
sliding code window.  However, the Montgomery multiplication kernel
has the interesting property shared by many real world codes that no
sequence of short superoptimizations will transform the code produced
by gcc -O3 into the code produced by STOKE.  We follow Bansal's approach
in overall system architecture, using testcases to help classify programs
as promising or not and eventually submitting the most promising candidates
to a verification engine to prove or refute their correctness.

More recently both Sketching
\cite{Solar-lezama06combinatorialsketching} and Brahma
\cite{DBLP:conf/pldi/GulwaniJTV11} have made progress in
addressing the closely related component-based program synthesis
problem.  These systems rely on either a declarative program
specification, or a user-specified partial program, and operate on
statements in bit-vector calculi rather than directly on hardware
instructions.  
Liang \cite{liang10programs} considers the task of learning programs 
from testcases alone, but at a similarly high level of abstraction.
Although useful for synthesizing results, the
internal representations used by these system preclude them from reasoning directly about the
runtime performance of the resulting code.

STOKE differs from previous approaches to superoptimization by relying on 
incomplete stochastic search.  In doing so, it makes heavy
use of Markov Chain Monte Carlo (MCMC) sampling to explore the extremely
high dimensional, irregular search space of loop-free assembly
programs.  For many optimization problems of this form, 
MCMC sampling is the only known general solution method which is also tractable.
Successful applications are many, and include protein alignment \cite{Neuwald97extractingprotein},
code breaking \cite{citeulike:6744178},
and scene modeling and rendering in computer graphics \cite{Veach97metropolislight, Chenney00samplingplausible}.

\section{Cost Minimization}
\label{sec:theory}

To cast program optimization as a cost minimization problem, 
  it is necessary to define a cost function with terms that balance the hard constraint of correctness preservation and the soft constraint of performance improvement.
The primary advantage of this approach is that it removes the burden of reasoning directly about the mutually-dependent optimization 
  issues faced by a traditional compiler.
For instance, rather than consider the interaction between register allocation and instruction selection, 
  we might simply define a term to encode the primary consequence: expected runtime.
Having done so, we may then utilize a cost minimization search procedure to discover a program that balances those issues as effectively as possible.
We simply run the procedure for as long as we like, and select the lowest-cost result which has satisfied all of the hard constraints.

In formalizing this idea, we make use of the following notation.
We refer to the input program as the {\em target} ($\mathcal{T}$) and a candidate compilation as a {\em rewrite} ($\mathcal{R}$),
  we say that a function $f(X;Y)$ takes inputs $X$ and is parameterized by $Y$,
  and finally, we define the indicator function for boolean variables: 
  
\begin{equation}
  {\bf 1}\{\phi\} = 
  \begin{cases}
    1 & \phi = {\rm true} \\
    0 & \phi = {\rm false} \\
  \end{cases}
\end{equation}

\subsection{Cost Function}

Although we have considerable freedom in defining a cost function, at the highest level, it should include two terms with the following properties:

\begin{equation}
  c(\mathcal{R}; \mathcal{T}) = \textrm{eq}(\mathcal{R}; \mathcal{T}) + \textrm{perf}(\mathcal{R}; \mathcal{T}) \\
\end{equation}
\begin{equation}
  \begin{array}{c}  
  {\rm eq}(\mathcal{R}; \mathcal{T}) = 0 \\
  \mathcal{R} = \arg \min_r \Big({\rm perf}(r; \mathcal{T)}\Big) \\
  \end{array}
\end{equation}

${\rm eq}(\cdot)$ is a correctness metric, measuring the similarity of
two functions.  The metric is zero if and only if the two functions
are equal.  For our purposes, two code sequences are regarded as
functions of registers and memory contents, and 
are are equal if for all machine states that agree on the
live inputs with the respect to the target, the two codes produce
identical side effects on the live outputs with respect to the target.
Because program optimization is undefined for ill-formed
programs, it is unnecessary that ${\rm eq}(\cdot)$ be defined for a
target or rewrite producing some undefined behavior.
However nothing prevents us from doing so, and it would be a straightforward extension to
produce a definition of ${\rm eq}(\cdot)$ which preserved hardware exception behavior as well.

${\rm perf}(\cdot)$ quantifies the performance improvement of a rewrite 
with respect to the target.
Depending on the application, this term could reflect code
size, expected runtime, number of disk accesses, power consumption, or
any other measure of resource usage.
Crucially, the extent to which this term accurately reflects the performance
improvement of a rewrite directly affects the quality of the
results discovered by a search procedure.

\subsection{MCMC Sampling}
\label{sec:mcmc}

In general, we expect cost functions of the form described above to be
highly irregular and not amenable to exact optimization techniques.
%It is highly unlikely that exist closed-form symbolic
%representations for either ${\rm eq}(\cdot)$ or ${\rm perf}(\cdot)$.
The common approach to solving this problem is to employ the use of an
MCMC sampler.  Although a complete discussion of MCMC 
is beyond the scope of this paper, we summarize the main ideas here.

MCMC is a technique for sampling from a probability density function in direct proportion to its value.
That is, regions of higher probability are sampled more often than regions of low probability.
When applied to cost minimization, it has the attractive property that in the limit
the most samples will be taken from the minimum (optimal) value of the function.
In practice, well before this limit behavior is observed, MCMC functions as an intelligent
hill climbing method which is robust against irregular functions that are dense with local minima.
A common method (described by \cite{citeulike:470600}) for transforming an arbitrary cost function, $c(\cdot)$, 
into a probability density function is the following,
where $\beta$ is a constant and $Z$ is a partition function that normalizes the
distribution:

\begin{equation}
  p(\mathcal{R}; \mathcal{T}) = \frac{1}{Z} \exp\Big(-\beta \cdot c(\mathcal{R}; \mathcal{T})\Big)
\end{equation}

Although computing $Z$ is in general intractable, the
Metro-polis-Hastings algorithm for generating Markov chains is
designed to explore density functions such as $p(\cdot)$ without the
need to compute the partition function \cite{annealing2,
  hastings1970montecarlo}.  The basic idea is simple.  The algorithm
maintains a current rewrite $\mathcal{R}$ and proposes a modified
rewrite $\mathcal{R}^*$ as the next step in the chain.  The {\em
  proposal} $\mathcal{R}^*$ is either accepted or rejected.  If the
proposal is accepted, $\mathcal{R}^*$ becomes the current rewrite,
otherwise another proposal based on $\mathcal{R}$ is generated.  The
algorithm iterates until its computational budget is exhausted, and so
long as the proposals are {\em ergodic} (capable of transforming any point
in the space to any other through some sequence of steps) the algorithm will in the
limit produce a sequence of samples with the properties described
above (i.e., in proportion to their cost).
This global property depends on the local acceptance criteria of a proposal $\mathcal{R} \rightarrow
\mathcal{R}^*$, which is governed by the Metropolis-Hastings acceptance
probability, where $q(\mathcal{R}^*|\mathcal{R})$ is the proposal
distribution from which a new rewrite $\mathcal{R}^*$ is sampled given
the current rewrite, $\mathcal{R}$:

\begin{equation}
  \alpha(\mathcal{R} \rightarrow \mathcal{R}^*; \mathcal{T}) = 
    \min \Bigg( 1, \frac{p(\mathcal{R}^*;\mathcal{T}) q(\mathcal{R}|\mathcal{R^*})}{p(\mathcal{R}; \mathcal{T}) q(\mathcal{R}^*|\mathcal{R})}\Bigg)
\end{equation}

This proposal distribution is key to a successful application of the algorithm. 
Empirically, the best results are obtained by a distribution which makes both local proposals that make minor modifications to $\mathcal{R}$
  and global proposals that induce major changes.
In the event that the proposal distributions are symmetric, $q(\mathcal{R}^*|\mathcal{R}) = q(\mathcal{R}|\mathcal{R}^*)$, 
  the acceptance probability can be reduced to the much simpler Metropolis ratio, 
  which can be computed directly from $c(\cdot)$:  

\begin{equation}
  \begin{split}
    \alpha(\mathcal{R} \rightarrow \mathcal{R}^*; \mathcal{T}) & = \min \Bigg(1, \frac{p(\mathcal{R}^*; \mathcal{T})}{p(\mathcal{R}; \mathcal{T})}\Bigg) \\
                                                               & = \min \Bigg(1, \exp\Bigg(-\beta \cdot \frac{c(\mathcal{R}^*; \mathcal{T})}{c(\mathcal{R}; \mathcal{T})}\Bigg) \Bigg) 
  \end{split}
  \label{eq:acc}
\end{equation}

The important properties of the acceptance criteria are the following:
%\begin{itemize}
%\item 
If $\mathcal{R}^*$ is better (has a higher probability/lower cost) than $\mathcal{R}$, the proposal is always accepted.
%\item 
If $\mathcal{R}^*$ is worse (has a lower probability/higher cost) than $\mathcal{R}$, the proposal may still be accepted
with a probability that decreases as a function of the ratio in value between $\mathcal{R}^*$ and $\mathcal{R}$.  
This is the property that prevents the search from becoming trapped in local minima while remaining less likely to accept a move that is 
much worse than available alternatives.
%\end{itemize}

\section{X86 Binary Optimization}
\label{sec:bin}

Having discussed program optimization as cost minimization in the abstract, 
  we turn to the practical details of implementing cost minimization for optimizing the runtime performance of 64-bit X86 binaries.
As 64-bit X86 is one of the most complex ISAs in production,
  we expect that the discussion in this section should generalize well to other architectures.

\subsection{Transformation Correctness}

\begin{figure}[t]
  \includegraphics[scale=0.65]{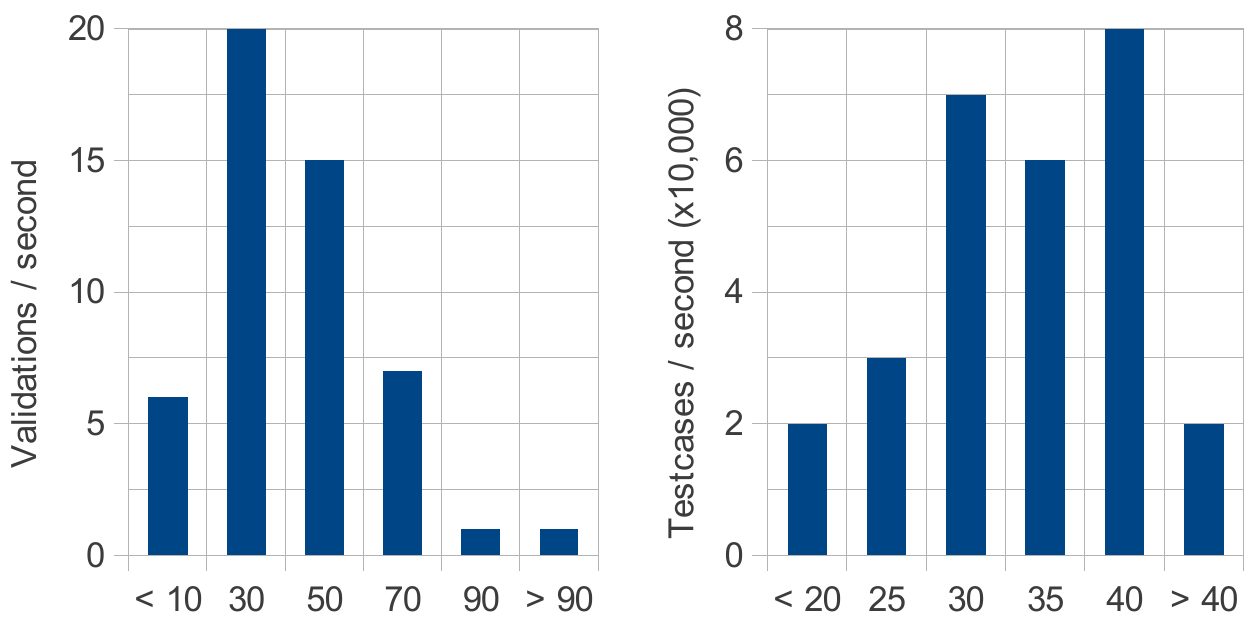}
  \caption{Histograms of validations per second (left), and testcase evaluations per second (right), for the benchmarks discussed in Section \ref{sec:eval}, 
           The low validation throughput is insufficient for MCMC.}
  \label{fig:validator}
\end{figure}  

For loop-free sequences of X86 assembly code, a natural choice for implementing the transformation correctness term 
  is a symbolic validator such as the one used in \cite{Cadar:2008:KUA:1855741.1855756}.
For a candidate rewrite, the term may be defined in terms of an invocation of the validator as:

\begin{equation}
  {\rm eq}(\mathcal{R}; \mathcal{T}) = 1 - \Big({\bf 1}\{{\rm VALIDATE}(\mathcal{T}, \mathcal{R})\}\Big)
  \label{eq:val}
\end{equation}

Unfortunately, despite advances in the technology, 
  the total number of validations that can be performed per
second, even for modestly sized codes, is low.  Figure
\ref{fig:validator} (left) suggests that for the 
benchmarks discussed in Section \ref{sec:eval} the number is well
below 100.  Because MCMC is
effective only insofar as it is able to explore sufficiently
large numbers of proposals, the repeated computation of Equation
\ref{eq:val} in its inner-most loop would almost certainly drive that
number well below a useful threshold.  

This observation motivates the
definition of an approximation to ${\rm eq}(\cdot)$ based on
testcases, $\tau$.  Intuitively, we run the proposal $\mathcal{R}^*$
on a set of inputs and measure ``how close'' the output is to the output
of the target on those same inputs.  For a given input, we use 
the number of bits difference in live outputs (i.e., the Hamming distance) to measure correctness.
Besides being much faster than using a theorem prover, this approximation of program
equivalence has the added advantage of producing a smoother landscape than the 0/1 output
of a symbolic equality test---it provides a useful notion of ``almost correct'' that can help guide
the search.

\begin{equation}
  \begin{split}
    {\rm eq}'(\mathcal{R}; \mathcal{T}, \tau) & = \sum_{t \in \tau} {\rm reg}(\mathcal{R}; \mathcal{T}, t) + {\rm mem}(\mathcal{R}; \mathcal{T}, t) \\
                                                 & +  \sum_{t \in \tau} {\rm err}(\mathcal{R}; \mathcal{T}, t)
  \end{split}                                                 
\end{equation}

${\rm reg}(\cdot)$ compares the side effects, ${\rm val}(\cdot)$, that both functions produce on live register outputs, $\rho$, with respect to the target,
  and counts the number of bits that the results differ by.
These outputs can include general purpose, SSE, and condition registers.
${\rm mem}(\cdot)$ is defined analogously for live memory outputs, $\mu$.
We use the population count function, ${\rm POP}(\cdot)$, to count the number
of 1-bits in the 64-bit representation of an integer.

\begin{equation}
  {\rm reg}(\mathcal{R}; \mathcal{T}, t) = \sum_{r \in \rho} {\rm POP}\Big({\rm val}(\mathcal{T}, r) \oplus {\rm val}(\mathcal{R}, r)\Big) \\
  \label{eq:reg}
\end{equation}
\begin{equation}
  {\rm mem}(\mathcal{R}; \mathcal{T}, t) = \sum_{m \in \mu} {\rm POP}\Big({\rm val}(\mathcal{T}, m) \oplus {\rm val}(\mathcal{R}, m)\Big)
\end{equation}

${\rm err}(\cdot)$ is used to distinguish programs which exhibit undefined behavior, by counting and then penalizing the number of 
  segfaults, ${\rm sigsegv}(\cdot)$, floating point exceptions, ${\rm sigfloat}(\cdot)$, and reads from undefined memory or registers, ${\rm undef}(\cdot)$,
  which occur during execution of the rewrite.
Note that ${\rm sigsegv}(\cdot)$ is defined in terms of the target, which determines the set of addresses which may be successfully dereferenced by a rewrite
  for a particular testcase.
Rewrites are run in a sandbox to ensure that undefined behavior can be detected safely.
The extension to additional kinds of counters would be straightforward.

\begin{equation}
  \begin{split}
    {\rm err}(\mathcal{R}; \mathcal{T}, t) &= w_{sf} \cdot {\rm sigsegv}(\mathcal{R}; \mathcal{T}, t) \\
                                           &+ w_{fp} \cdot {\rm sigfloat}(\mathcal{R}; t) \\
                                           &+ w_{ur} \cdot {\rm undef}(\mathcal{R}; t)
  \end{split}
\end{equation}

The evaluation of ${\rm eq'}(\cdot)$ may be accomplished either by JIT
compilation, or the use of a hardware emulator.  For this paper we
have chosen the latter.  Figure \ref{fig:validator}(right) shows
the number of testcase executions that our emulator is able to perform
per second: just under 500,000.  This implementation allows us to
define an optimized method for computing ${\rm eq}(\cdot)$ which
achieves sufficient throughput to be useful for MCMC.

\begin{equation}
{\rm eq}^*(\mathcal{R}; \mathcal{T}, \tau) = 
  \begin{cases}
    {\rm eq}(\mathcal{R}; \mathcal{T})      & {\rm eq}'(\mathcal{R}; \mathcal{T}, \tau) = 0 \\
    {\rm eq}'(\mathcal{R}; \mathcal{T}, \tau) & {\rm otherwise}
  \end{cases}
  \label{eq:opt_val}
\end{equation}

In addition to performance, Equation \ref{eq:opt_val} has the following desirable properties.
First, failed computations of ${\rm eq}(\cdot)$ will produce a counterexample testcase that may be used to refine $\tau$ as described in \cite{Cadar:2008:KUA:1855741.1855756}.
The careful reader will note that refining $\tau$ affects the cost function, $c(\cdot)$, and effectively changes the search space that it defines.
However in practice, the number of failed validations that are required to produce a robust set of testcases that accurately predict success is quite low.
Second, as discussed above, it smooths the search space by allowing the transformation equality metric to quantify how different two codes are.

\subsection{Performance Improvement}
\label{sec:perf}

\begin{figure}[t]
  \centering
  \includegraphics[scale=0.65]{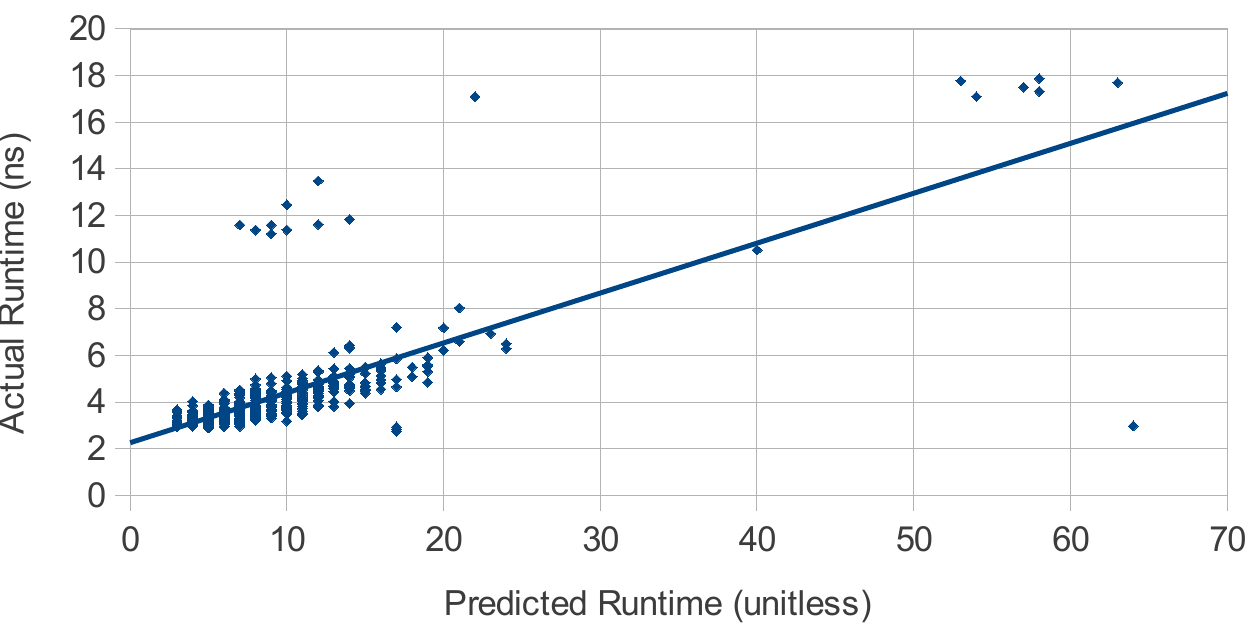}
  \caption{Comparison of predicted and actual runtimes for the benchmarks described in Section \ref{sec:eval}, along with rewrites generated while writing this paper.
           The points are well correlated but distinguished by outliers corresponding to codes with high instruction level parallelism at the micro-op level.
           The approximation is sufficient for the benchmarks we consider.} 
  \label{fig:optimization}
\end{figure}  

A straightforward method for computing the performance improvement term is to JIT compile both the target and the rewrite code and compare their runtimes.
Unfortunately, as with the transformation correctness term, the amount of time required to both compile a function and execute it sufficiently many times to eliminate
  transient performance effects is prohibitively expensive to be used in MCMC's inner-most loop.
For this paper, we adopt a simple heuristic for approximating the runtime performance of a function, 
  which is based on a static approximation of the average latency of its instructions.

\begin{equation} 
  \begin{split}
  {\rm perf}(\mathcal{R}; \mathcal{T}) & = H(\mathcal{T}) - H(\mathcal{R}) \\
  H(f)                                 & = \sum_{i \in {\rm inst}(f)} {\rm LATENCY}(i) 
  \end{split}
\end{equation}

Figure \ref{fig:optimization} shows a high correlation between the
heuristic and the actual runtimes of the 
benchmarks described in Section \ref{sec:eval}, along with rewrites
for those benchmarks which were generated in the process of writing
this paper.  Outliers correspond to rewrites with a disproportionately
high or low amount of instruction level parallelism at the micro-op level.  
A more accurate model of the second order performance effects introduced by a modern
CISC processor is straightforward if tedious to construct and we
expect would be necessary for some programs.  However, the approximation
is largely sufficient for the benchmarks we consider in this paper.

Small errors of this form can be addressed by recomputing ${\rm perf}(\cdot)$ using the slower JIT compilation method as a postprocessing step.
We simply record the top-n lowest cost samples taken by MCMC, rerank them based on their actual runtimes, and return the best result.

\subsection{MCMC Sampling}

For X86 binary optimization, candidate rewrites are finite loop-free sequences of instructions, of length $\ell$,
  where a distinguished token, ${\rm UNUSED}$, allows for the representation of programs with fewer than $\ell$ instructions.
This simplifying assumption is essential to the formulation of MCMC discussed in Section \ref{sec:mcmc}, 
  as it places a constant value on the dimensionality of the search space.
The interested reader may consult \cite{citeulike:1327191} for a thorough treatment of why this is necessary.
Our definition of the proposal distribution, $q(\cdot)$, chooses among four possible moves: the first two minor, and the latter two major:

{\bf Opcode.} 
  With probability $p_c$,
  an instruction is selected at random, and its opcode is replaced by a random opcode.
  The new opcode is drawn from an equivalence class of opcodes expecting the same number and type of operands as the old opcode.
  For this paper, we construct these classes from the set of arithmetic and fixed point SSE opcodes.

{\bf Operand.} 
  With probability $p_o$,
  an instruction is selected at random and one of its operands is replaced by a random operand
  drawn from an equivalence class of operands with types equivalent to the old operand.
  If the operand is an immediate, its value is drawn from a bag of predefined constants.

{\bf Swap.} 
  With probability $p_s$,
  two instructions are selected at random and interchanged.

{\bf Instruction.}  With probability $p_i$, an instruction is selected
at random, and its opcode is replaced either by an unconstrained
random instruction or the ${\rm UNUSED}$ token.  A random instruction
is constructed by first selecting an opcode at random and then
choosing random operands of the appropriate types.  The ${\rm UNUSED}$
token is proposed with probability $p_u$.

These definitions satisfy the ergodicity property described in Section \ref{sec:mcmc}.
Any program can be transformed into any other through repeated application of Instruction moves.
These definitions also satisfy the symmetry property, and thus allow the computation of acceptance probability using Equation \ref{eq:acc}.
To see why, note that the probabilities of performing all four moves types are equal to the probabilities of undoing the transformations they produce using a move of the same type.
The opcode and operand moves are constrained to sample from identical equivalence classes before and after acceptance.
Similarly, the swap and instruction moves are equally unconstrained in both directions.

\subsection{Separating Synthesis From Optimization}

\begin{figure}[t]
  \centering
  \includegraphics[scale=1.0]{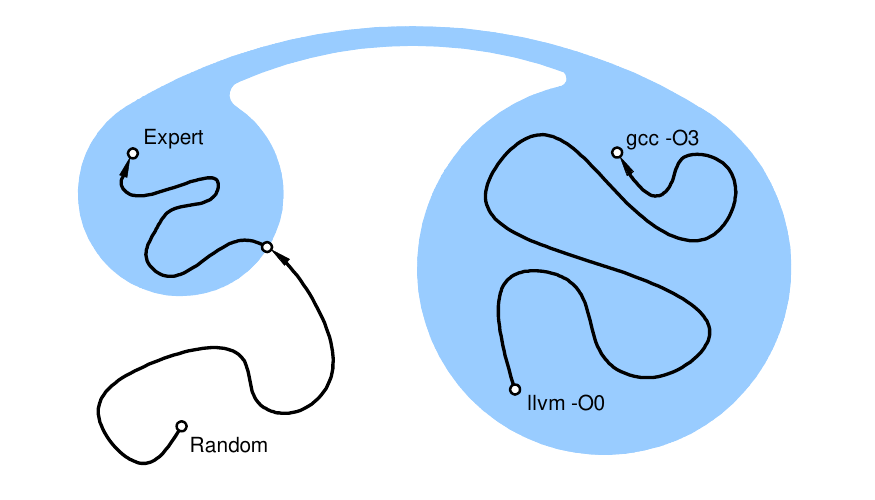}
  \caption{Abstract depiction of the search space for the Montgomery multiplication benchmark.
           O0 and O3 optimized codes occupy a densely connected part of the space which is easily traversed.
           Expert code occupies an entirely different region of the space reachable only by an extremely low probability path.}

  \label{fig:search_space}
\end{figure}  

An early implementation of STOKE, based on the above principles, was
able to consistently transform llvm -O0 code into the equivalent of gcc
-O3 code.  Unfortunately, it was rarely able to produce code
competitive with expert hand-written code.  The reason is suggested by
Figure \ref{fig:search_space}, which gives an abstract depiction of the
search space for the Montgomery multiplication benchmark.  For
loop-free sequences of code, llvm -O0 and gcc -O3 codes differ
primarily with respect to efficient use of the stack and choices of
individual instructions.  Yet despite these differences, the resulting
codes are algorithmically quite similar.  To see this, note that compiler
optimizers are generally designed to compose many small local transformations:
dead code elimination deletes one instruction, constant propagation changes one
register to an immediate, strength reduction replaces a multiplication with an add.
With respect to the search space, such sequences of local optimizations occupy a region 
of equivalent programs that are densely connected by very short sequences of moves (often
a single move) that is easily traversed by a local search method.  Beginning from llvm
-O0 code, a random search method will quickly identify local
inefficiencies one by one, improve them in turn, and hill climb its
way to a gcc -O3 code.

The expert code discovered by STOKE occupies an entirely different
region of the search space.  As noted earlier, it has the property
that no sequence of small equality preserving transformations connect
it to either the llvm -O0 or the gcc -O3 code.  It represents a
completely distinct algorithm for implementing the Montgomery
multiplication kernel at the assembly level.  The only method we know
of for a local search procedure to transform either code
into the expert code is to traverse the extremely low probability
path that builds the expert code in place next to the original, all the
while increasing its cost, only to delete the original code at the very end.  Although
MCMC is guaranteed to traverse this path in the limit, the
likelihood of it doing so in any reasonable amount of time is so low as to be useless in practice.

This observation motivates dividing the cost minimization into two phases:
\begin{itemize}
\item 
A {\em synthesis phase} focused solely on correctness, which attempts to locate regions of equal programs 
   distinct from the region occupied by the target.
\item 
An {\em optimization phase} focused on speed, which searches for the fastest program within each of those regions.
\end{itemize}

The two phases share the same search implementation; only the starting point and the acceptance functions are different.
Synthesis begins with a random starting point (a sequence of randomly chosen instructions), while optimization begins with
a code sequence known to be equivalent to the target.
For proposals, synthesis ignores the performance improvement term altogether and simply uses
  Equation \ref{eq:opt_val} as its cost function.
Optimization uses both terms, allowing it to measure improvement while also allowing it to experiment with ``shortcuts'' that
(temporarily) violate transformation correctness.

\subsection{Optimized Acceptance Computation}

\begin{figure}[t]
  \centering
  \includegraphics[scale=0.65]{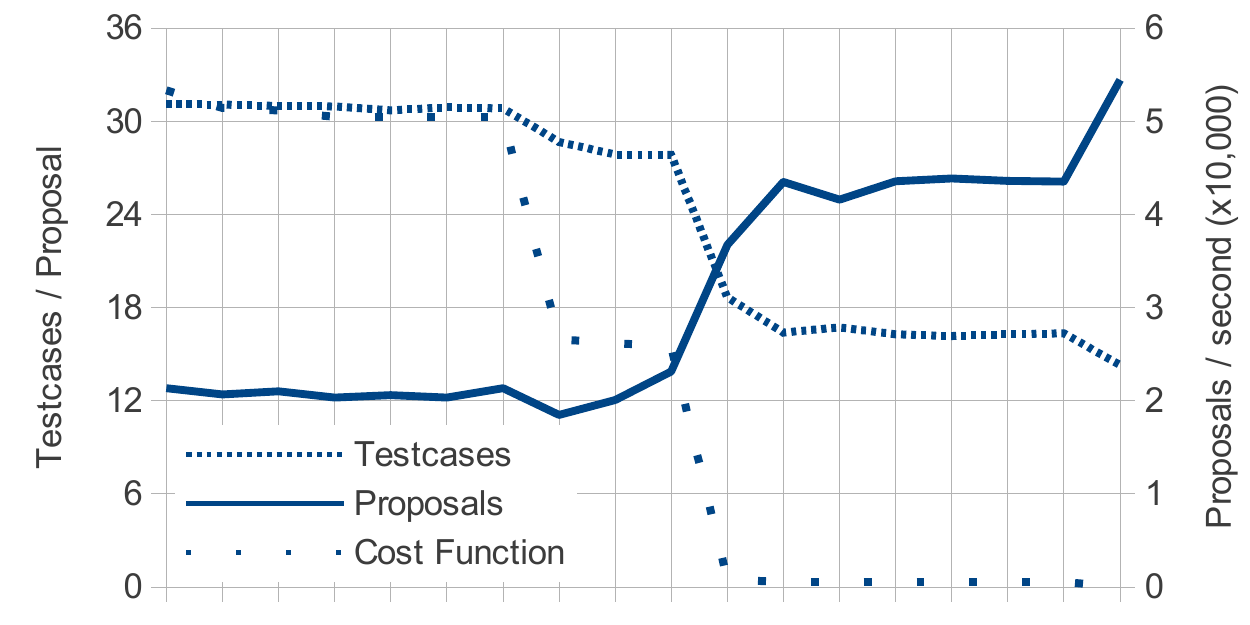}
  \caption{Proposals evaluated per second versus testcases evaluated prior to early termination, 
           during synthesis for the Montgomery multiplication benchmark.
           Reducing the number of evaluated testcases produces an almost 3x improvement in proposal throughput.
           Cost function shown unitless.}
  \label{fig:tester}
\end{figure}  

\begin{figure}[t]
  \begin{minipage}{\linewidth}
    \centering
    \includegraphics[scale=0.8]{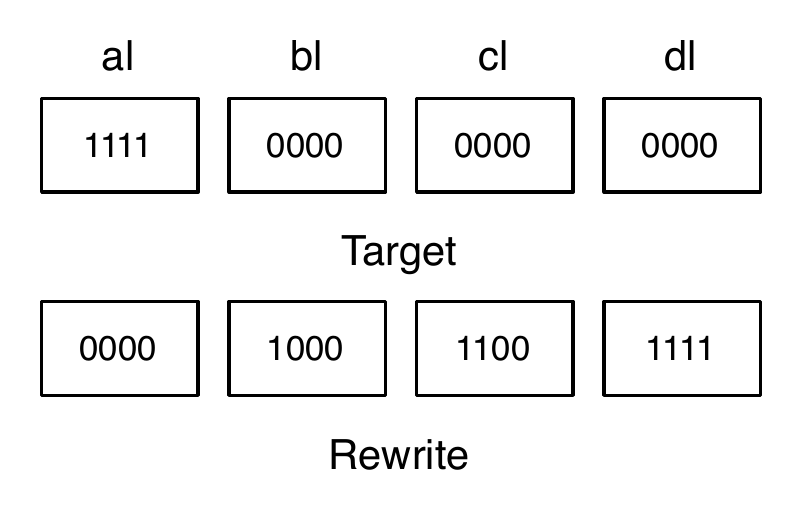}
  \end{minipage}
  \begin{minipage}{\linewidth}
    \centering
    \begin{tabular}{r|c|c|c|c}
                                                                                                   & {\bf al} & {\bf bl} & {\bf cl} & {\bf dl} \\
       ${\rm val}(\mathcal{T}, {\rm al}) \oplus {\rm val}(\mathcal{R}, \cdot)$                     & 1111     & 0111     & 0011     & 0000     \\
       ${\rm POP}\Big({\rm val}(\mathcal{T}, {\rm al}) \oplus {\rm val}(\mathcal{R}, \cdot)\Big)$  & 4        & 3        & 2        & 0        \\
       $w_m \cdot {\bf 1}\{{\rm al} \neq \cdot\}$                                                  & 0        & 1        & 1        & 1        \\
    \end{tabular}
  \end{minipage}
  \begin{minipage}{\linewidth}
    \begin{displaymath}
      \begin{split}
        {\rm reg}(\mathcal{T}, \mathcal{R}, \tau)  &= 4 \\
        {\rm reg}'(\mathcal{T}, \mathcal{R}, \tau) &= \min(4, 3+1, 2+1, 1) \\
                                                   &= 1 \\
      \end{split}
    \end{displaymath}
  \end{minipage}
  \caption{Strict versus improved equality functions for a machine state in which ${\rm ax}$ is live out.
           Strict assigns the maximum possible cost to a rewrite which produces the correct value in the wrong location.
           Improved assigns a cost of almost zero.}
  \label{fig:relaxed}
\end{figure}  

\begin{figure}[t]
  \centering
  \includegraphics[scale=0.65]{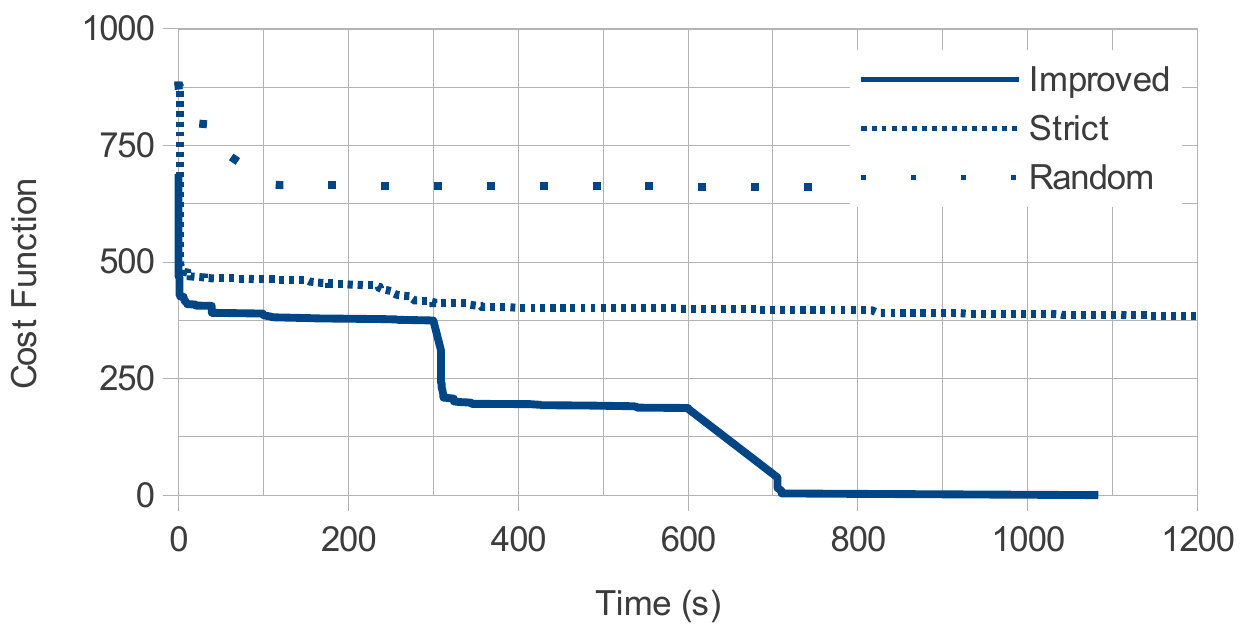}
  \caption{Strict versus improved synthesis cost functions for the Montgomery multiplication benchmark.
           In the amount of time (s) required for improved to converge, strict produces a result similar to a purely random search.
          }
  \label{fig:synthesis_fxn}
\end{figure}

The optimized method for computing ${\rm eq}^*(\cdot)$ given in
Equation \ref{eq:opt_val} is sufficiently fast for MCMC.
However, its performance can be further improved.  As described so
far, ${\rm eq}^*(\cdot)$ is computed by first running the proposal
on the testcases, summing their costs, noting the ratio in total cost
with the current rewrite, and then sampling a random variable to decide
whether to accept the proposal.  Instead, we can sample the random variable $p$
first, compute the maximum value of the ratio we can accept given $p$,
and then run testcases but terminate early if the bound is exceeded.

More technically, because our formulation of the proposal distribution
$q(\cdot)$ is symmetric we may compute the acceptance probability
$\alpha(\cdot)$ of a proposal directly from $c(\cdot)$ as shown in
Equation \ref{eq:acc}.  By first sampling 
$p$ we can invert $\alpha(\cdot)$ to solve for the
maximum cost rewrite $c(\cdot)$ that we will accept.

\begin{equation}
  \begin{split}
    p                                   & < \alpha(\mathcal{R} \rightarrow \mathcal{R}^*; \mathcal{T}) \\
                                        & < \min \Bigg(1, \exp\Bigg(-\beta \cdot \frac{c(\mathcal{R}^*; \mathcal{T})}{c(\mathcal{R}; \mathcal{T})}\Bigg) \Bigg) \\
    c(\mathcal{R}^*; \mathcal{T}, \tau) & < c(\mathcal{R}; \mathcal{T}, \tau) - \frac{\log(p)}{\beta}  
  \end{split}
\end{equation}

Because the computation of ${\rm eq}'(\cdot)$ is based on the
iterative evaluation of testcases, it is only necessary to do so for as long
as the running sum does not exceed this upper bound.  Once it does, we
know that the proposal is guaranteed to be rejected, and no further
computation is necessary.  Figure \ref{fig:tester} shows the result of
applying this optimization during synthesis for the
Montgomery multiplication benchmark.  As the
value of the cost function decreases, so too do the average number of
testcases which must be evaluated prior to early termination.  This in
turn produces a considerable increase in the number of testcases
evaluated per second, which at peak exceeds 50,000.

\subsection{Improved Equality Metric}
\label{sec:imp_eq}

A second and even more important improvement stems from the
observation that the definition of ${\rm reg}(\cdot)$ given in Equation
\ref{eq:reg} is unnecessarily strict.  Figure \ref{fig:relaxed} gives
an illustrative example.  Consider a machine with
four 4-bit registers, and a target function that produces side
effects on register ${\rm al}$.  The final machine states produced by
running the target and a candidate rewrite are shown at the top of the
figure.  Because the value that the rewrite produces for ${\rm al}$ 
has no correct bits the rewrite is assigned
the maximum possible cost.  However the rewrite does produce
the correct value, only in the wrong location: ${\rm dl}$.  The improvement
is to reward rewrites that produce correct (or nearly correct) values in
the wrong places.   The improved cost function
examines all registers of equivalent
bit-width ${\rm bw}(\cdot)$ and selects the one that matches the target register most closely, assigning
an additional small penalty if the selected register is not the correct one:  
\begin{equation}
  \begin{split}
    {\rm reg}'(\mathcal{R}; \mathcal{T}, \tau) &= \sum_{r \in \rho} \min_{r' \in {\rm bw}(r)} {\rm R}(r, r'; \tau) \\
    {\rm R}(r, r'; \tau)                       &= {\rm POP}\Big({\rm val}(\mathcal{T}, r) \oplus {\rm val}(\mathcal{R}, r')\Big) \\
                                               &+ w_m \cdot {\bf 1}\{r \neq r'\} 
  \end{split}
\end{equation}
For
brevity, we note that we improve the definition of memory equality 
analogously.

Figure \ref{fig:synthesis_fxn} shows the results of using the improved
definitions of register and memory equality during synthesis for the
Montgomery multiplication benchmark.  In the amount of
time required for the improved cost function to converge to a
zero-cost rewrite, the strict version obtained a minimum cost which
was only slightly superior to that obtained by a pure random search.
The dramatic increase in performance can be explained as an implicit
parallelization of the search procedure.  By allowing a candidate
rewrite to place a correct value in an arbitrary location, the
improved cost function allows candidate rewrites to simultaneously
explore as many alternate computations as can be fit within a sequence
of length $\ell$.

\subsection{Why and When Synthesis Works}
\label{sec:why}

\begin{figure}[t]
  \centering
  \includegraphics[scale=0.65]{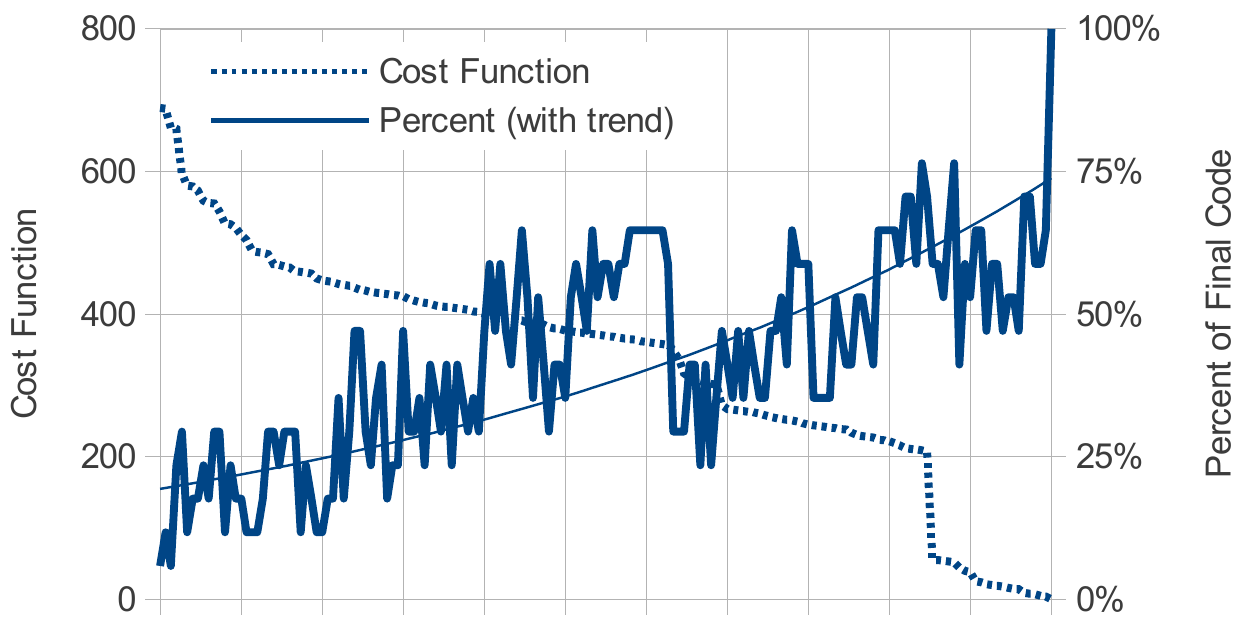}
  \caption{Cost function versus percentage of instructions which appear relative to final zero-cost rewrite.
           Random search is an effective method for performing synthesis insofar as it is able to discover partially correct rewrites incrementally.}
  \label{fig:synthesis}
\end{figure}  

\begin{figure*}[t]
  \centering
  \includegraphics[scale=0.8]{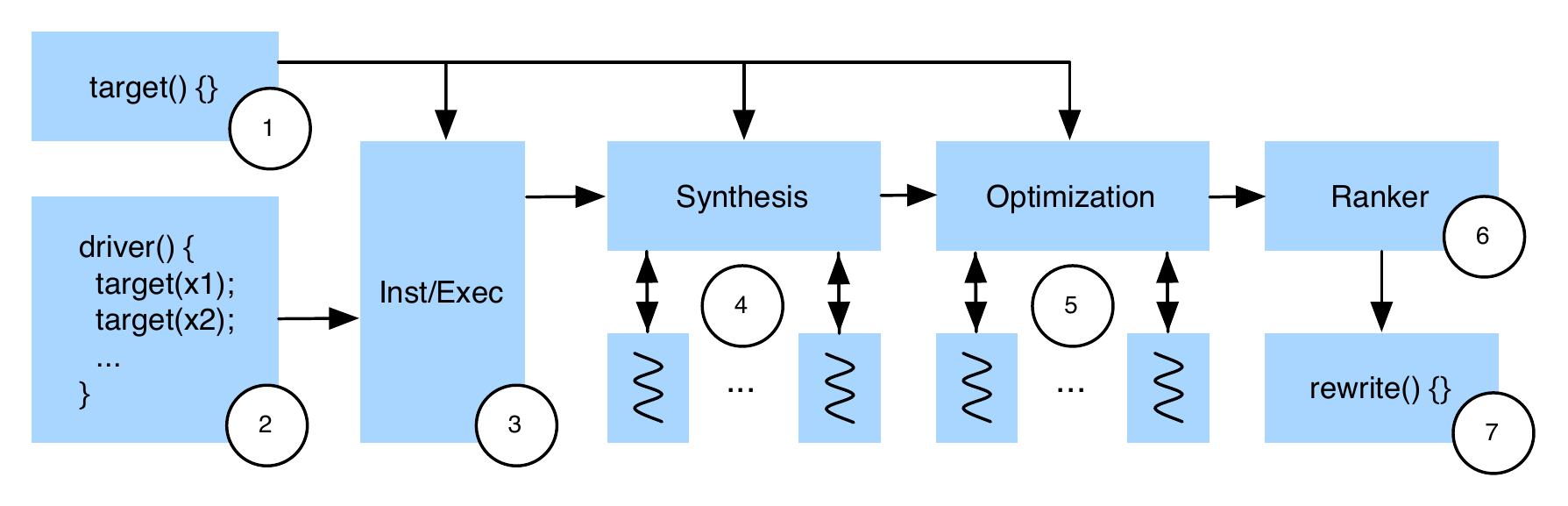}
  \caption{The high-level design of STOKE.  A target binary created by a production compiler (1) and driver code (2) are run under instrumentation (3) using automatically generated inputs to produce testcases.
           Synthesis threads (4) use the target and testcases to generate candidate rewrites, which along with the target are refined by optimization threads (5).
           The results are ranked (6) and the rewrite with the lowest cost is returned to the user (7).}
  \label{fig:system}
\end{figure*}  

It is not intuitive that a randomized search procedure should synthesize
a correct rewrite from such an enormous search space in a short
amount of time.  In our experience, the secret is that synthesis is effective
precisely when it is possible to discover parts of a correct
rewrite incrementally, as opposed to all at once.  Figure
\ref{fig:synthesis} plots the current best cost obtained
during synthesis against the percentage of instructions appearing
in both that rewrite and the final correct rewrite for the Montgomery
multiplication benchmark.  As search proceeds, the percentage of
correct code  increases in inverse
proportion to the value of the cost function.  While this is
very encouraging and there are many programs that satisfy the property that
they can be synthesized in pieces, each of which increases the average number of correct bits in the output,
there are certainly interesting programs that do not have
this property.  In the limit, any code performing a complex
computation that is reduced to a single boolean value poses
a problem for our approach.  The discovery of
partially correct computations is useful as a guide for random search
only insofar as they are able to produce a partially correct result,
which can be detected by a cost function.
      
This observation motivates the desire for a cost function which
maximizes the signal produced by a partially correct rewrite.  We
discussed a successful application of this principle in Section
\ref{sec:imp_eq}.  Nonetheless, there remains room for improvement.
Consider the program which rounds its inputs up to the next highest
power of two.  This program has the interesting property that it
differs from the program which simply returns zero in only one bit
per testcase.  The improved cost function discussed above assigns
a very low cost to the constant zero function, which although almost correct is
completely wrong, and exhibits no partially correct computations that
can be hill-climbed to a correct rewrite.

Fortunately, we note that even when synthesis fails, optimization is still possible.
It must simply proceed only from the region occupied by the target as a starting point.

\section{STOKE}
\label{sec:sys}

STOKE is a prototype implementation of the concepts described in this paper with high-level design shown in Figure \ref{fig:system}.
A user provides a target binary which was created using a production compiler (in our case, llvm -O0);
  in the event that the target contains loops, STOKE identifies loop-free subsequences of the code which it will attempt to optimize.
The user also provides an annotated driver in which the target is called in an appropriate context.
Based on the user's annotations, STOKE automatically generates random inputs to the target, compiles the driver, and then runs the code under instrumentation to produce testcases.
The target and testcases are broadcast to a small cluster of synthesis threads which after a fixed amount of time report back candidate rewrites.
In like fashion, a small cluster performs optimization on both the target and those rewrites.
Finally, the set of rewrites with a final cost that is within 20\% of the minimum are re-ranked based on actual runtime, and the best is returned to the user.

\subsection{Test Case Generation and Evaluation}

\begin{figure*}[t]
  \centering
  \includegraphics[scale=0.7]{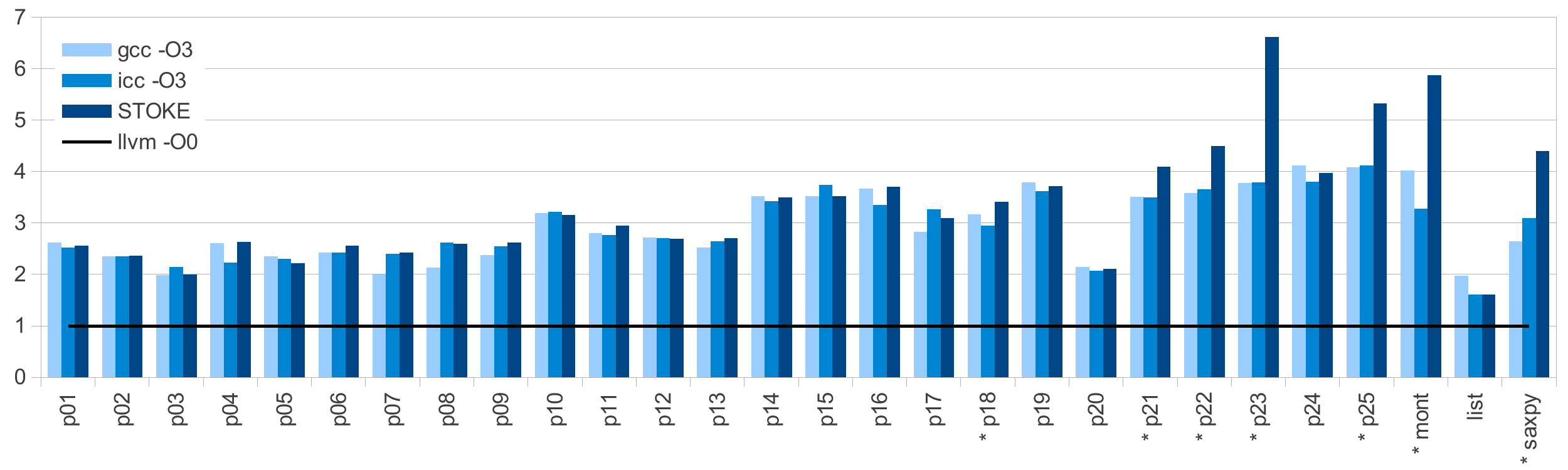}
  \caption{Average speedup over llvm -O0 for benchmark kernels.  
           Beginning from code produced by llvm -O0, STOKE discovers rewrites which are comparable to code produced by gcc and icc with full optimizations enabled.
           In some cases, the rewrite outperforms both, and are comparable to expert handwritten assembly.
           Kernels for which STOKE discovered an algorithmically distinct rewrite are annotated with a star.}
  \label{fig:runtime}
\end{figure*}  

STOKE automatically generates testcases using annotations provided by a user.
Because STOKE operates on 64-bit X86 assembly, those inputs are limited to fixed-width bit strings, which unless otherwise specified, are sampled uniformly at random.
If the target uses an input to form a memory address, the user must annotate that input with a range of values that guarantee that the resulting addresses are legal given the context in which the target is called.
The compiled program is executed under instrumentation using Intel's PinTool \cite{Luk05pin:building}.  
As each instruction is executed, the tool records the state of all general purpose, SSE, and condition registers, as well as dereferenced memory.
The initial state of the registers, along with the first values dereferenced from each memory address are used to form testcase inputs.
Outputs are formed analogously.
By default, STOKE generates 32 testcases for each target.

For each testcase, The set of addresses dereferenced by the target are used to define the sandbox in which candidate rewrites are executed.
Attempts to dereference invalid addresses are trapped and replaced by instructions which produce a constant zero value.
Attempts to read from registers in an undefined state and computations which produce floating point exceptions are handled similarly.

\subsection{Validation}

STOKE uses a sound procedure for validating the equality of two sequences of loop-free assembly which is similar to the one described in \cite{DBLP:conf/asplos/BansalA06}.
Code sequences are converted into SMT formulae in the quantifier free theory of bit-vector arithmetic used by the STP \cite{stp} theorem prover,
  and used to produce a query which asks whether both sequences produce the same side effects on live outputs when executed from the same initial machine state.
For our purposes, a machine state consists of general purpose, SSE, and condition registers, and memory.
Depending on type, registers are modeled as between 8- and 128-bit vectors.
Memory is modeled as two vectors: a 64-bit address and an 8-bit value (X86 is byte addressable).

\begin{figure}[t]
  \centering
  \begin{tabular}{||c|c||c|c||c|c||}
    \hline
      $w_{sf}$ & 1 & $p_c$ & 0.16 & $p_u$   & 0.16 \\
    \hline
      $w_{fp}$ & 1 & $p_o$ & 0.5  & $\beta$ & 0.1  \\
    \hline
      $w_{ur}$ & 2 & $p_s$ & 0.16 & $\ell$  & 50   \\
    \hline
      $w_{m}$  & 3 & $p_i$ & 0.16 &         &      \\
    \hline
  \end{tabular}
  \caption{MCMC parameters used by STOKE for synthesis and optimization.}
  \label{fig:mcmc_param}
\end{figure}

\begin{figure*}[t]
  \centering
  \includegraphics[scale=0.7]{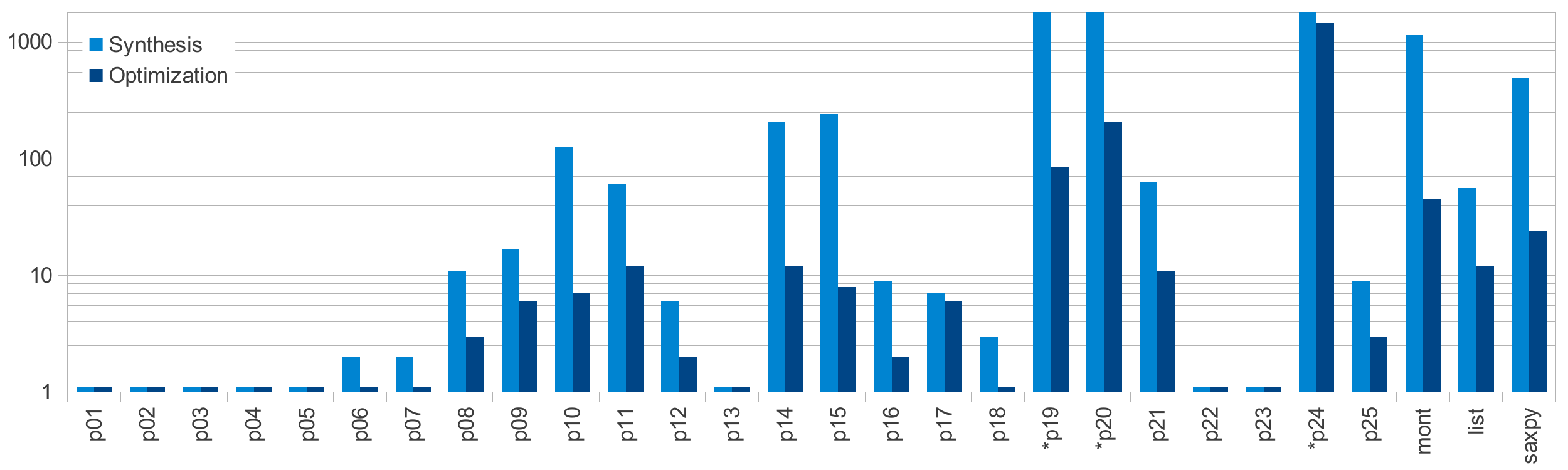}
  \caption{STOKE runtimes for synthesis and optimization (s) required to produce the results shown in Figure \ref{fig:runtime}.
           Kernels for which synthesis timed out are annotated with a star.}
  \label{fig:timing}
\end{figure*}  

STOKE first asserts the constraint that both sequences agree on the initial machine state of the live inputs with respect to the target.
Next, it iterates over the instructions in the target, and for each instruction asserts a constraint which encodes the transformation it produces on the machine state.
These constraints are chained together to produce a constraint on the final machine state of the live outputs with respect to the target.
Analogous constraints are asserted for the rewrite.
Finally, for all pairs of memory accesses at addresses ${\tt addr_1}$ and ${\tt addr_2}$, STOKE asserts an additional constraint which relates their values:
${\tt addr_1}={\tt addr_2}\Rightarrow {\tt val_1}={\tt val_2}$.
Using these constraints, STOKE performs an STP query which asks whether there does not exist an initial machine state which causes
  the two sequences to produce different values for the live outputs with respect to the target.
If the answer is ``yes", then the sequences are equal.
If the answer is ``no", then the prover produces a counter example which is used to produce a new testcase.

STOKE makes two simplifying assumptions which are necessary to keep validator runtimes tractable.
First, it assumes that stack addresses are represented exclusively as constant offsets from the stack pointer.
This allows STOKE to treat stack addresses as nameable locations, and minimizes the number of expensive memory constraints which must be asserted.
This is essential for validating against llvm -O0 code, which exhibits heavy stack traffic.
Second, it treats 64-bit multiplication and division as uninterpreted functions, by asserting the constraint that the instructions
  produce identical random values when executed on identical inputs.
Whereas STP diverges when reasoning explicitly about two or more such operations, our benchmarks contain as many as four per sequence.

\subsection{Parallel Synthesis and Optimization}

Synthesis and optimization are executed in parallel on a small cluster consisting of 40
  dual-core 1.8 GHz AMD Opterons.
Both are allocated computational budgets of 30 minutes.
The MCMC parameters used by both phases are summarized in Figure \ref{fig:mcmc_param}.

\section{Evaluation}
\label{sec:eval}

In addition to the Montgomery multiplication kernel discussed
so far, STOKE was evaluated on benchmarks drawn both from 
literature and real-world high-performance codes.  
The performance improvements obtained for those kernels are summarized in Figure \ref{fig:runtime}, while
corresponding STOKE runtimes are shown in Figure \ref{fig:timing}.
Beginning with a binary compiled by llvm -O0, STOKE consistently discovers rewrites
which match the performance of the code produced by gcc and icc with full optimizations enabled.
In several cases, the performance exceeds both and is comparable to expert handwritten assembly.
As we explain below, the improvement often results from the discovery
of a completely distinct assembly level algorithm for implementing the target code.
We close with discussion of the benchmarks which highlight STOKE's limitations.

\subsection{Hacker's Delight}

\begin{figure}[t]
    \tt
\begin{lstlisting}
int p21(int x, int a, int b, int c) {
  return ((-(x == c)) & (a ^ c)) ^
         ((-(x == a)) & (b ^ c)) ^ c;
}
\end{lstlisting}
\begin{parcolumns}[rulebetween=true]{2}
\colchunk[1]{\begin{acol}
.L0                   
  movl  edx, eax
  xorl  edx, edx
  xorl  ecx, eax
  cmpl  esi, edi
  sete  dl
  negl  edx
  andl  edx, eax
  xorl  edx, edx
  xorl  ecx, eax
  cmpl  ecx, edi
  sete  dl
  xorl  ecx, esi
  negl  edx
  andl  esi, edx
  xorl  edx, eax
\end{acol}}
\colchunk[2]{\begin{acol}
.L0
  cmpl   edi, ecx
  cmovel esi, ecx
  xorl   edi, esi
  cmovel edx, ecx
  movq   rcx, rax
\end{acol}}
\end{parcolumns}
\rm
  \centering
  \caption{Cycling Through 3 Values benchmark.
           STOKE sees through the esoteric implementation which gcc -O3 translates literally (left) and rediscovers the intuitive algorithm using conditional move intrinsics (right).}
  \label{fig:p18}
\end{figure}

Hacker's Delight \cite{Warren:2002:HD:515297}, commonly referred to as ``the bible of bit-twiddling hacks", is a collection of techniques
  for encoding otherwise complex algorithms as small loop-free sequences of bit-manipulating instructions.
Gulwani \cite{DBLP:conf/pldi/GulwaniJTV11} noted this as a fine source of benchmarks for program synthesis and superotpimization,
  and identified a 25 program benchmark which ranges in complexity from turning off the right-most bit in a word, 
  to rounding up to the next highest power of 2, or selecting the upper 32 bits from a 64-bit multiplication.
Our implementation of the benchmark uses the C code found in the original text.
For brevity, we discuss only the programs for which STOKE discovered an algorithmically distinct rewrite.

Figure \ref{fig:p18} shows the ``Cycle Through 3 Values" benchmark, which takes an input, ${\rm x}$, and transforms it to the next value in the sequence $\langle {\rm a}, {\rm b}, {\rm c} \rangle$: ${\rm a}$ becomes ${\rm b}$, 
  ${\rm b}$ becomes ${\rm c}$, and ${\rm c}$ becomes ${\rm a}$.
Hacker's Delight points out that the most natural implementation of this function is a sequence of conditional assignments,
  but notes that on an ISA without conditional move intrinsics the implementation shown is cheaper than one which uses branch instructions.
For 64-bit X86, which has conditional move intrinsics, this turns out to be an instance of premature optimization.
Unfortunately, neither gcc nor icc are able to detect this, and are forced to transcribe the code as written.
There are no sub-optimal subsequences in the resulting code and the production compilers are simply unable to reason about the semantics of the function as a whole.
For this reason, we expect that equality-preserving superoptimizers would exhibit the same behavior.
STOKE on the other hand, has no trouble rediscovering the natural implementation from the 41 line llvm -O0 compilation.
We note that although this rewrite is only five lines long, it remains beyond the reach of superoptimizers based on bruteforce enumeration.

In similar fashion, the implementation that Hacker's Delight recommends for the ``Compute the Higher Order Half of a 64-bit Product" multiplies two 32-bit inputs in four parts and aggregates the results.
The computation resembles the Montgomery multiplication benchmark, and STOKE discovers a rewrite which requires a single multiplication using the appropriate bit-width intrinsic.
STOKE additionally discovers a number of typical superoptimizer rewrites.
These include using the ${\rm popcnt}$ intrinsic, which counts the number of 1-bits in an integer, as an intermediate step in the 
  ``Compute Parity" and ``Determine if an Integer is a Power of 2" benchmarks.

\subsection{SAXPY}

\begin{figure}[t]
    \tt
  \begin{minipage}{\linewidth}
  \begin{lstlisting}
  void SAXPY(int* x, int* y, int a) {
    x[i]   = a * x[i]   + y[i];
    x[i+1] = a * x[i+1] + y[i+1];
    x[i+2] = a * x[i+2] + y[i+2];
    x[i+3] = a * x[i+3] + y[i+3];
  }
  \end{lstlisting}
  \end{minipage}
  \begin{minipage}{\linewidth}
  \begin{lstlisting}
  .L0                   
    movslq ecx, rcx       
    leaq   (rsi,rcx,4), r8  
    leaq   1(rcx),      r9       
    movl   (r8),        eax        
    imull  edi,         eax        
    addl   (rdx,rcx,4), eax 
    movl   eax,         (r8)        
    leaq   (rsi,r9,4),  r8
    movl   (r8),        eax 
    imull  edi,         eax
    addl   (rdx,r9,4),  eax
    leaq   2(rcx),      r9
    addq   3,           rcx
    movl   eax,         (r8)
    leaq   (rsi,r9,4),  r8
    movl   (r8),        eax
    imull  edi,         eax
    addl   (rdx,r9,4),  eax
    movl   eax,         (r8)
    leaq   (rsi,rcx,4), rax
    imull  (rax),       edi
    addl   (rdx,rcx,4), edi
    movl   edi,         (rax)

  .L0
    movd   edi,         xmm0
    shufps 0,           xmm0, xmm0
    movups (rsi,rcx,4), xmm1
    pmullw xmm1,        xmm0
    movups (rdx,rcx,4), xmm1
    paddw  xmm1,        xmm0
    movups xmm0,        (rsi,rcx,4)
  \end{lstlisting}
  \end{minipage}
  \rm
  \centering
  \caption{SAXPY benchmark.
           Unlike gcc -O3 (top), STOKE discovers a rewrite which uses SSE vector instructions (bottom).}
  \label{fig:saxpy}
\end{figure}

SAXPY (Single-precision Alpha X Plus Y) is a level 1 vector operation in the Basic Linear Algebra Subsystems Library \cite{Blackford01anupdated}.
The code makes heavy use of heap accesses and presents the opportunity for optimization using vector intrinsics.
To enable STOKE to discover this possibility, our implementation is unrolled four times by hand, as shown in Figure \ref{fig:saxpy}.
Despite heavy annotation to indicate that the addresses pointed to by $x$ and $y$ are aligned and do not alias each other,
  the production compilers either cannot detect the possibility of a compilation using vector intrinsics,
  or are precluded by some internal heuristic from doing so. 
STOKE on the other hand, discovers the natural implementation:
  the constant $a$ is broadcast four ways from a general purpose register into an SSE register, 
  and then multiplied by, and added to the contents of $x$ and $y$, which are loaded into SSE registers four elements at a time.
The four way broadcast does not appear anywhere in either the gcc -O3 code, or in the original 61 line llvm -O0 code.
As observed above, this and the length of the final rewrite allow STOKE to outperform both the production compilers and likely existing superoptimizers as well.

\subsection{Limitations}

\begin{figure}[t]
    \tt
\begin{lstlisting}
while ( head != 0 ) {
  head->val *= 2;
  head = head->next;
}
\end{lstlisting}
\begin{parcolumns}[rulebetween]{2}
    \colchunk[1]{\begin{acol}
 movq -8(rsp), rdi
.L4               
 sall  (rdi)      
 movq  8(rdi), rdi
.L6               
 testq rdi,    rdi
 jne   .L4        
\end{acol}}
\colchunk[2]{\begin{acol}
.L4                     
 movq  -8(rsp), rdi    
 sall  (rdi)           
 movq  8(rdi), rdi    
 movq  rdi, -8(rsp)
.L6                     
 movq  -8(rsp), rdi
 testq rdi, rdi    
 jne   .L4             
\end{acol}} \end{parcolumns}
\rm
  \centering
  \caption{Linked List Traversal benchmark.
           STOKE discovers the same rewrite (right) as Bansal's superoptimizer,
           but fails to cache the head pointer in a register, as in the gcc -O3 code (left).}
  \label{fig:bansal}
\end{figure}

\begin{comment}
The Montgomery multiplication rewrite shown in Figure \ref{fig:teaser} is one of several in an equivalence class of rewrites which are all assigned the minimum possible cost by our performance heuristic.
Within that equivalence class, performance varies between 1.2 and 1.6x improvement over the code produced by gcc -O3 (Figure \ref{fig:runtime} reports the best of those values).
The reason for the variation is the micro-op instruction level parallelism discussed in Section \ref{sec:perf}, which our performance heuristic is unable to account for.
The effect appears in the P14 Hacker's Delight benchmark as well, and accounts for the poor runtime.
A more accurate model of the second order effects produced by a modern processor would account for this and allow us to neatly reify the instruction scheduling phase of a traditional compiler into our framework.
\end{comment}

Bansal's superoptimizer \cite{DBLP:conf/asplos/BansalA06} was evaluated on the Linked List Traversal Benchmark shown in Figure \ref{fig:bansal}.
The code iterates over a list of integers and multiplies each of the elements by two.
The code is unique with respect to the benchmarks discussed so far, as it contains a loop.
As a result, STOKE is unable to optimize the function as a whole, but rather only it's inner-most loop-free fragment.
STOKE discovers the same optimizations as Bansal's superoptimizer, the elimination of stack traffic and a strength reduction from multiplication to bit shifting.
However it fails in like fashion to eliminate the instructions which copy the head pointer from and to the stack on every iteration of the loop.
The production compilers on the other hand, are able to eliminate the memory traffic by caching the pointer in a register prior to entering the loop.
As a result, the rewrite discovered by STOKE is slower than the code produced by gcc -O3 (surprisingly, icc does not perform strength reduction, and produces code which performs similarly).
This shortcoming could be addressed by extending our framework to validate and propose modifications to code containing loops. 

As shown in Figure \ref{fig:timing}, STOKE is unable to synthesize a rewrite for three of the Hacker's Delight Benchmarks.
All three benchmarks, despite being quite complex, have the interesting property that they produce results which differ by only a single bit from a simple yet completely incorrect alternative.
The ``Round Up to the Next Highest Power of 2" benchmark is nearly indistinguishable from the function which always returns zero.
The same is true of the ``Next Highest with Same Number of 1-bits", and a small transformation to the ``Exchanging Two Fields" benchmark with respect to the identity function.
Fortunately, for these three benchmarks, using its optimization phase alone STOKE is still able to discover rewrites which perform comparably to the production compiler code, which we believe to be optimal.
In general, however, we do not expect this to be the case.
A more sophisticated cost function, as described in section \ref{sec:why}, is surely necessary.

\section{Conclusion and Future Work}

We have shown a new approach to the loop-free binary superoptimization task which reformulates program optimization as a stochastic search problem.
Compared to a traditional compiler, which factors optimization into a sequence of small independently solvable subproblems,
  our framework is based on cost minimization and considers the competing constraints of transformation correctness and performance improvement simultaneously as terms in a cost function.
We show that an MCMC sampler can be used to rapidly explore functions of this form and produce low cost samples which correspond to high quality code sequences.
Although our method sacrifices completeness, the scope of programs which we are able to reason about, 
  and the quality of the rewrites we produce, far exceed those of existing superoptimizers.

Although our prototype implementation, STOKE, is in many cases able to produce rewrites which are competitive with or outperfrom the code produced by production compilers,
  there remains substantial room for improvement.
In future work, we intend to pursue both
  a validation and proposal mechanism for code containing loops 
  and a synthesis cost function which is robust against targets with numerous deceptively attractive, albeit completely incorrect synthesis alternatives.

\section{Acknowledgments}

The authors would like to thank Peter Johnston, Juan Manuel Tamayo, and Kushal Tayal for their assistance in the implementation of STOKE,
Ankur Taly for his help with semantics of X86 opcodes, David Ramos for his help with STP, 
  Jake Herczeg for his time spent designing Figure \ref{fig:search_space}, 
  and Martin Rinard for suggesting a technique which focuses more on testcases than validation.

\bibliographystyle{abbrv}
\bibliography{stoke}  % sigproc.bib is the name of the Bibliography in this case
\end{document}